\documentclass[12pt]{article}
\usepackage{amsfonts,amsmath,amssymb}
\usepackage{scalerel}[2016/12/29]
\usepackage{graphicx}
\usepackage{tikz}

\def\be{\begin{equation}}
\def\ee{\end{equation}}
\def\bea{\begin{eqnarray}}
\def\eea{\end{eqnarray}}

\usepackage{amsmath}
\usepackage{epsfig,graphicx}
\usepackage{graphicx}
\usepackage{tikz}

\def\be{\begin{equation}}
\def\ee{\end{equation}}
\def\bea{\begin{eqnarray}}
\def\eea{\end{eqnarray}}

\begin{document}

\thispagestyle{plain}

\title{\bf\Large Exclusion statistics and lattice random walks}

\author{St\'ephane Ouvry$^*$ \  {\scaleobj{0.9}{\rm and}} Alexios P. Polychronakos$^\dagger$}


\maketitle

\begin{abstract}
We establish a connection between exclusion statistics with arbitrary integer exclusion parameter $g$ and a class of random walks on planar lattices. This connection maps the generating function for the number of closed walks of given length enclosing 
a given algebraic area on the lattice to the grand partition function of particles obeying exclusion statistics $g$ in a particular
single-particle spectrum, determined by the properties of the random walk. Square lattice random walks, described in terms of the Hofstadter Hamiltonian, correspond to $g=2$. In the $g=3$ case we explicitly construct a corresponding chiral random walk model on a triangular lattice, and we point to potential random walk models for higher $g$. In this context,
we also derive the form of the microscopic cluster coefficients for arbitrary exclusion statistics.
\end{abstract}

\noindent
* LPTMS, CNRS,  Universit\'e Paris-Sud, Universit\'e Paris-Saclay,\\ \indent 91405 Orsay Cedex, France; {\it stephane.ouvry@u-psud.fr}

\noindent
$\dagger$ Department of Physics, City College of New York, NY 10038, USA; \\ \indent
{\it apolychronakos@ccny.cuny.edu}
\vskip 1cm

\vfill
\eject

\section{Introduction}%
\label{sec1}

The enumeration of closed random walks of a given length with a given
algebraic area on a planar lattice is a challenging subject. Algebraic
area is defined as the oriented area spanned by the walk as it traces
the lattice. A unit lattice cell enclosed in the counterclockwise
(positive) way has an area $+1$, whereas when enclosed in the clockwise
(negative) way it has an area $-1$. The~total algebraic area is the area
enclosed by the walk weighted by the winding number: if the walk winds
around more than once, the area is counted with multiplicity.

It is well known that the algebraic area enumeration can be mapped to
the quantum problem of a particle hopping on the lattice pierced by a
perpendicular homogeneous magnetic field. Indeed, in quantum mechanics,
the magnetic field is coupled to the area spanned by the particle. In
\cite{nous} a closed formula for the enumeration on a square
lattice was proposed. The relevant quantum model in that case is the
celebrated Hofstadter model \cite{Hofstadter} of a particle
hopping on a square lattice in a perpendicular magnetic field.

The enumeration of closed walks of (necessarily even) length
$\mathbf{n}$ for the square lattice was achieved by studying the secular
determinant of the Hofstadter Hamiltonian. This
determinant was calculated in \cite{Kreft} for a rational flux per
plaquette in units of the elementary flux quantum, in which case the
determinant becomes finite dimensional. The coefficients of the
expansion of this determinant in powers of the energy, called Kreft
coefficients, are given by certain multiple nested trigonometric
sums that are reminiscent of partition functions. The expression
of the area enumeration generating function for the lattice walks of
a given length was derived \cite{nous} in terms of a
different set of coefficients, extracted from the Kreft coefficients,
themselves reminiscent of cluster coefficients. These facts hinted to
an interpretation in terms of statistical mechanics of many-body systems.

Motivated by these observations, we revisit the enumeration of closed
lattice random walks enclosing a given algebraic area and demonstrate
that it does admit a statistical mechanical interpretation, but in terms
of particles obeying generalized \emph{exclusion statistics} with
exclusion parameter $g$ ($g=0$ for bosons, $g=1$ for fermions and higher
$g$ means a stronger exclusion beyond Fermi).

Exclusion statistics was proposed by Haldane \cite{Haldane} as a
distillation of the statistical mechanics properties of Calogero-like
spin systems. The relation of Calogero particles and fractional
statistics was first pointed out in \cite{NRBos}. Exclusion
statistics also emerges in the context of anyons projected on the lowest
Landau level of a strong magnetic field \cite{Das}, and has
further been extended to other situations \cite{Poly}. It is a
remarkable fact that algebraic area considerations in lattice walks also
end up amounting to particular many body systems with exclusion
statistics.

One payoff of the uncovered connection is that, once it is established
from its defining properties that a random walk corresponds to specific
exclusion statistics, the expression for the number of walks with given
algebraic area can be straightforwardly extracted from statistical
mechanical expressions. In this context, square lattice walks correspond
to $g=2$, where the Kreft coefficients appear as many-body partition
functions of exclusion-2 particles and the algebraic area generating
functions as the corresponding cluster coefficients. We will also give
an explicit construction of triangular lattice walks realizing $g=3$
statistics and hint at other generalized walks corresponding to
statistics with higher values of the exclusion parameter.

The algebraic area generating function of a lattice walk is determined
by the exclusion statistics parameter $g$ as well as a quantity (which
we call the spectral function) derived from the properties of the walk.
We will demonstrate that walks of different statistics and spectral
functions can nevertheless be equivalent, a ``fermionization'' result
mapping systems with different spectra and statistics.

Finally, an additional bonus of our analysis is an explicit expression
for the cluster coefficients of particles with exclusion statistics
$g$ in a collection of single-particle quantum states with arbitrary
energies. These coefficients were derived before in the thermodynamic
limit, but their exact expressions for a set of microscopic states
(i.e., \emph{not} in the thermodynamic limit) were not known~\cite{Ouv}.

\section{Lattice algebraic area enumeration and the Hofstadter model: a review}%
\label{sec2}

We will start by presenting a review of the Hofstadter model and the
results in \cite{nous} in order to fix ideas and establish
notations.

Consider closed random walks of length $\mathbf{{n}}$ on a square
lattice ($\mathbf{{n}}=2, 4, 6, \ldots $ is necessarily even) and denote
by $C_{\mathbf{{n}}}(A)$ the number of such walks enclosing an algebraic
area $A$ ($A$ is between $-\lfloor (\mathbf{{n}}/4)^{2}\rfloor $ and
$\lfloor (\mathbf{{n}}/4)^{2}\rfloor $ where $\lfloor \;\rfloor $
denotes the `floor' function). Obviously $C_{\mathbf{{n}}}(A)=C_{
\mathbf{{n}}}(-A)$.

The enumeration $C_{\mathbf{{n}}}(A)$ is achieved by establishing a
relation between random lattice walks and the Hofstadter system of a
charged particle hopping on a square lattice with a magnetic flux per
unit cell $\phi $. The Hofstadter Hamiltonian is
%
\begin{equation}
H=u+u^{-1}+v+v^{-1}
\label{Hof}%
\end{equation}
where $u$ and $v$ are respectively the hopping operators on the
horizontal and vertical axis. Denoting $\mathrm{Q}=\exp (2 i \pi
\phi /\phi _{0})$, with $\phi _{0}$ the unit of flux quantum, the hopping
operators obey the ``noncommutative torus'' relation
%
\begin{equation}
\label{com}
v\; u = \mathrm{Q}\; u\; v
\end{equation}
due to the noncommutativity of magnetic translations when a flux is
piercing the lattice. The connection to random walks and their algebraic
area $A$ is established through
\begin{equation}
\nonumber
v^{-1}\; u^{-1} \; v\; u=\mathrm{Q} ~.
\end{equation}
The operators (in the sequence they act on the right) represent a walk
enclosing one elementary square cell in the counterclockwise (positive)
sense. Clearly the power of $\mathrm{Q}$ represents the enclosed area,
and the above relation generalizes to products of $u,v,u^{-1},v^{-1}$
representing arbitrary (closed) walks on the lattice, their algebraic
area appearing as $\mathrm{Q}^{A}$. The generating function for closed
walks of length $\mathbf{n}$ is then given by $\mathrm{Tr}\:H^{
\mathbf{{n}}}$ under an appropriate normalization of the trace such that
$\mathrm{Tr}\:1 = 1$.

When the flux per square $\phi $ is a rational multiple of the flux
quantum $\phi _{0}$, that is, $\phi /\phi _{0}=p/q$ with $p$ and $q$
co-prime, $\mathrm{Q}=\exp (2 i \pi p/q)$, the representation of $u$ and $v$ becomes finite dimensional
%
\begin{equation}
u=e^{ik_{y} }
\begin{pmatrix}
\mathrm{Q}& 0 & 0 & \cdots & 0 & 0
\\
0 & \mathrm{Q}^{2} &0& \cdots & 0 & 0
\\
0 & 0 & \mathrm{Q}^{3} & \cdots & 0 & 0
\\
\vdots & \vdots & \vdots & \ddots & \vdots & \vdots
\\
0 & 0 & 0 & \cdots &\mathrm{Q}^{q-1} & 0
\\
0 & 0 & 0 & \cdots & 0 & \mathrm{Q}^{q}
\\
\end{pmatrix}
\label{umat}%
\end{equation}
%
\begin{equation}
v=
\begin{pmatrix}
0
\;\;
& 1
\;\,
& 0
\;\;
& 0\cdots
\;\;
& 0
\;\;
& 0\;
\\
0
\;\;
& 0
\;\;
&1
\;\;
& 0\cdots
\;\;
& 0
\;\;
& 0\;
\\
0
\;\;
& 0
\;\;
& 0
\;\;
&1 \cdots
\;\;
& 0
\;\;
& 0\;
\\
\vdots
\;\;
& \vdots
\;\;
& \vdots
\;\;
& \ddots
\;\;
& \vdots
\;\;
& \vdots \;
\\
0
\;\;
& 0
\;\;
& 0
\;\;
&0 \cdots
\;\;
& 0
\;\;
& 1 \;
\\
e^{i q k_{x} }
\;\;
& 0
\;\;
& 0
\;\;
&0
\;\;
\cdots & 0
\;\;
& 0\;
\\
\end{pmatrix}
\label{vmat}%
\end{equation}
where $k_{x}$ and $k_{y}$ are the quasimomenta in the $x$ and $y$
directions. $u$ and $v$ satisfy the commutation relation (\ref{com}).
The operators
%
\begin{equation}
u^{q}=e^{i q k_{y}} ~,
~~~
v^{q}=e^{i q k_{x}}
\end{equation}
commute with $u$ and $v$ and constitute Casimirs of the algebra, their
values in the above representation determined by the quasimomenta.

In the\vadjust{\vspace{1pt}} quantum problem, $k_{x}$ and $k_{y}$ span the range ($-\frac{
\pi }{q},\frac{\pi }{q}$) and produce the Hofstadter energy bands. In
the mapping to lattice walks the full trace involves also integrating
over $k_{x} , k_{y}$, and this results in the elimination of terms
containing $u^{q}$ and $v^{q}$ which do not correspond to closed walks
but contribute to $\mathrm{tr} H^{\mathbf{n}}$. We will call such
unwanted terms ``umklapp'' effects. A simpler approach is to simply set
$k_{x} = k_{y} = 0$ and consider walks of length less than $q$, or
otherwise remove the above umklapp terms. In the following we will use
{(\ref{Hof})} and $u$, $v$ in {\eqref{umat}}, {\eqref{vmat}} with
$k_{x} = k_{y} = 0$ and still call this simplified Hamiltonian the
``Hofstadter Hamiltonian''.

The Hofstadter spectrum stems from the secular determinant of the
$q \times q$ matrix
%
\begin{eqnarray}
\nonumber
\det (1-z H)&=&
\begin{vmatrix}
1-z(\mathrm{Q}+\frac{1}{\mathrm{Q}}) & -z & 0 & \cdots & 0 & -z
\\
-z & 1-z (\mathrm{Q}^{2}+\frac{1}{\mathrm{Q}^{2}}) &-z & \cdots & 0 &
0
\\
0 & -z & () & \cdots & 0 & 0
\\
\vdots & \vdots & \vdots & \ddots & \vdots & \vdots
\\
0 & 0 & 0 & \cdots & () & -z
\\
-z & 0 & 0 & \cdots & -z & 1-z(\mathrm{Q}^{q}+
\frac{1}{\mathrm{Q}^{q}})
\\
\end{vmatrix}
\\
&=&-\sum _{j=0}^{\lfloor q/2 \rfloor }a_{p, q}(2j)z^{2j}-4z^{q}~.
\label{matrix}%
\end{eqnarray}

The $a_{p,q}(2j)$ are the so-called Kreft coefficients
\cite{Kreft},
%
\begin{eqnarray}[ll]
\nonumber
a_{p,q}(2j)
\\
\nonumber
\quad=(-1)^{j+1}\sum _{k_{1}=0}^{q-2j}\sum _{k_{2}=0}^{k_{1}}
\ldots \sum _{k_{j}=0}^{k_{j-1}}
 4\sin ^{2}\left (\frac{\pi (k_{1}+2j-1)
p}{q}\right )4\sin ^{2}\left (\frac{\pi (k_{2}+2j-3) p}{q}\right )
\ldots
\\
\quad
\phantom{\quad=(-1)^{j+1}\sum _{k_{1}=0}^{q-2j}\sum _{k_{2}=0}^{k_{1}}
\ldots \sum _{k_{j}=0}^{k_{j-1}}} 4\sin ^{2}\left (\frac{\pi (k_{j-1}+3) p}{q}\right )4\sin ^{2}\left (\frac{
\pi (k_{j}+1) p}{q}\right )
\label{00}
\end{eqnarray}
with $a_{p,q}(0)=-1$, while $-4 z^{q}$ in (\ref{matrix}) is an umklapp term.

By scaling, the Kreft coefficient $a_{p,q}({2j})=q [q] a_{p,q}({2j})+
\ldots +q^{{j}}[q^{{j}}]a_{p,q}({2j})$ turns out to be a polynomial in
$q$ of order $j$ with the $k$th order coefficients $[q^{k}] a_{p,q}(
{2j})$, $1\le k\le j$, being linear combinations of the $\cos ({2A
\pi p/q})$'s.

The enumeration of closed lattice walks of length $\mathbf{n}$ with
algebraic area $A$ is possible \cite{nous} since the generating
function for the $C_{\mathbf{{n}}}(A)$'s coincides with the
$1$st order (i.e. proportional to $q$) term in the above polynomial,
which represents $\mathrm{Tr}\:H^{\mathbf{{n}}}$
%
\begin{equation}
\label{simplify}
\frac{1}{{\mathbf{{n}}}}\sum _{A} C_{\mathbf{{n}}}(A) \mathrm{Q}^{A }
=[q] a_{p,q}(\mathbf{{n}})~.
\end{equation}

The Kreft coefficients appear in the secular determinant of the
Hofstadter Hamiltonian and therefore can be expressed in terms of traces
of powers of this Hamiltonian. As a consequence, higher order terms in
the $q$ expansion of $a_{p,q}(\mathbf{{n}})$ are also given in terms of
the linear terms of lower-order coefficients $ [q]a_{p,q}(2n)$,
$n\le {\mathbf{{n}}}/2$. E.g.,
%
\begin{eqnarray}
\nonumber
a_{p,q}(2)
&=& q[q]a_{p,q}(2)
\\
\nonumber
a_{p,q}(4)
&=&q[q]a_{p,q}(4)-\frac{q^{2}}{2!}\left ([q]a_{p,q}(2)\right )
^{2}
\\
\nonumber
a_{p,q}(6)
&=&q[q]a_{p,q}(6)-q^{2}[q]a_{p,q}(2)[q]a_{p,q}(4)+\frac{q
^{3}}{3!}\left ([q]a_{p,q}(2)\right )^{3}
\nonumber
\\
\nonumber
a_{p,q}(8)
&=&q[q]a_{p,q}(8)-q^{2}[q]a_{p,q}(2)[q]a_{p,q}(6)-\frac{q
^{2}}{2!}\left ([q]a_{p,q}(4)\right )^{2}
\\
\nonumber
&+&\frac{q^{3}}{2!}\left ([q]a_{p,q}(2)\right )^{2}[q]a_{p,q}(4)-\frac{q
^{4}}{4!}\left ([q]a_{p,q}(2)\right )^{4}
\\
{\mathrm{etc.}}\;,
&
\label{cluster}
\end{eqnarray}
i.e.,
%
\begin{equation}
a_{p,q}(\mathbf{{n}})=\, -
\!\sum _{
{k_{n}\ge 0 \atop \sum _{n} nk_{n}= \mathbf{{n}}/2}} \prod _{n=1}
^{\mathbf{{n}}/2}(-1)^{k_{n}} \frac{1}{k_{n} !}\left (q[q]a_{p,q}(2n)\right )
^{k_{n}}~.
\label{000}%
\end{equation}

Denoting
%
\begin{equation}
4 \sin ^{2}(\pi k p/q)={{{\tilde{b}}}_{p/q}}(k)
\label{denote}%
\end{equation}
$[q]a_{p,q}(\mathbf{{n}})$ can be expressed as a
linear combination of the building blocks
\begin{equation}
\sum _{k=1}^{q} {{{\tilde{b}}}_{p/q}}^{{l_{1}}}(k){{{\tilde{b}}}_{p/q}}
^{l_{2}}(k-1)\ldots {{{\tilde{b}}}_{p/q}}^{l_{j}}(k-j+1)
\nonumber
\end{equation}
labeled by the compositions\footnote{There are $2^{\mathbf{{n}}/2-1}$
such compositions; for example, when $\mathbf{{n}}=8$, one has the
$8$ compositions $8/2=4=3+1=1+3=2+2=2+1+1=1+2+1=1+1+2=1+1+1+1$.} of
$\mathbf{{n}}/2$, that is, the ordered set of any number of positive
integers $(l_{1},\dots ,l_{j} )$ satisfying $\mathbf{{n}}/2=l_{1}+l
_{2}+\ldots +l_{j}$, $1\le j\le {\mathbf{{n}}}/2$, namely
%
\begin{equation}
[q]a_{p,q}(\mathbf{{n}})=
\sum _{
{l_{1}, l_{2}, \ldots , l_{j} \atop { \mathrm{composition}}\;\mathrm{of}\;\mathbf{{n}}/2}}
c(l_{1},l_{2},\ldots ,l_{j} ) \, \frac{1}{q}\sum _{k=1}
^{q} {{{\tilde{b}}}_{p/q}}^{{l_{1}}}(k){{{\tilde{b}}}_{p/q}}^{l_{2}}(k-1)
\ldots {{{\tilde{b}}}_{p/q}}^{l_{j}}(k-j+1)
\label{label}%
\end{equation}
where
%
\begin{equation}
\label{ouf}
c(l_{1},l_{2},\ldots ,l_{j})= \frac{\binom{l_{1}+l_{2}}{l_{1}}}{l_{1}+l
_{2}}
\;\;
l_{2}\frac{\binom{l_{2}+l_{3}}{l_{2}}}{l_{2}+l_{3}}\;\ldots
\;\;
l_{{j}-1}\frac{\binom{l_{{j}-1}+l_{j}}{l_{{j}-1}}}{l_{{j}-1}+l_{j}}
\end{equation}
and
%
\begin{equation}
\sum _{
{l_{1}, l_{2}, \ldots , l_{j} \atop {\mathrm{composition}}\;\mathrm{of}\;\mathbf{{n}}/2}}
 c(l_{1},l_{2},\ldots ,l_{j})=\frac{\binom{\mathbf{{n}}}{
{\mathbf{{n}}/2}}}{\mathbf{{n}}}\;.
\label{right}%
\end{equation}
On the other hand \cite{nous}, ${{{\tilde{b}}}_{p/q}}(k)$ in
{(\ref{denote})} is such that the coefficients in the $\cos ({2 A
\pi p/q})$ expansion of $(1/q)\sum _{k=1}^{q} {{{\tilde{b}}}_{p/q}}^{{l_{1}}}(k)
{{{\tilde{b}}}_{p/q}}^{l_{2}}(k-1)\ldots {{{\tilde{b}}}_{p/q}}^{l_{j}}(k-
{j}+1)$'s also add up to $\binom{\mathbf{{n}} }{{\mathbf{{n}}}/2}$ for
any composition $l_{1},l_{2},\ldots ,l_{j}$ of $\mathbf{{n}}/2$. These
two factors concur to give the number of closed random walks of (even)
length $\mathbf{{n}}$ to be
\begin{equation*}
\binom{\mathbf{{n}} }{{\mathbf{{n}}}/2}^{2},
\end{equation*}
as it should. Note that this overall counting can directly be retrieved
by taking the limit $q \to \infty $ so that $\mathrm{Q}\to 1$. In that
limit all the ${\tilde{b}}_{p/q}(k)$ factors in each building block
become equal, so the sum in {(\ref{label})} factorizes into two sums
\begin{equation}
[q]a_{p,q}(\mathbf{{n}})=
\sum _{
{l_{1}, l_{2}, \ldots , l_{j} \atop { \mathrm{composition}}\;\mathrm{of}\;\mathbf{{n}}/2}}
c(l_{1},l_{2},\ldots ,l_{j} ) ~
\frac{1}{q}\sum
_{k=1}^{q} {{{\tilde{b}}}_{p/q}}^{{\mathbf{{n}}/2}}(k)
\nonumber
\end{equation}
and the second sum goes over to the integral
\begin{equation}
\nonumber
\frac{1}{q}\sum _{k=1}^{q} {{{\tilde{b}}}_{p/q}}^{{\mathbf{{n}}/2}}(k)
\to \int _{0}^{1} {{{\tilde{b}}}_{p/q}}^{{\mathbf{{n}}/2}}(q s)\; ds =
\int _{0}^{1} (2 \sin {\pi p s})^{\mathbf{n}}\; ds = \binom{\mathbf{n}
}{{\mathbf{n}}/2}
\end{equation}
reproducing the second factor $\binom{\mathbf{{n}}}{{\mathbf{{n}}/2}}$.

We conclude by noting that the expressions in {(\ref{cluster})} and
{(\ref{000})} are essentially cluster expansions of $a_{p,q}(
\mathbf{{n}})$ viewed as a partition function in terms of the
$[q]a_{p,q}(2n)$'s, $n\le {\mathbf{{n}}}/2$, which then play the role
of cluster coefficients. This is the first hint that there is a
statistical mechanical interpretation at hand. In the next section we
will make this correspondence explicit.

\section{Lattice random walks and $g=2$ exclusion statistics}%
\label{sec3}
Our key observation is that, looking at {(\ref{cluster})} and {(\ref{000})},
the area generating function for random walks $[q]a_{p/q}(
\mathbf{{n}})$ in {(\ref{simplify})} can be viewed as the cluster
coefficient for the partition function $a_{p/q}(\mathbf{{n}})$ of a
system of particles with exclusion statistics of order $g=2$. To this
end, let us interpret $\tilde{b}_{p/q}(k)$ as a spectral function
\begin{equation}
\tilde{{b}}_{p/q}(k) := \exp (-\beta \epsilon _{k})\;,
\nonumber
\end{equation}
then $a_{p/q}(2j)$ in {(\ref{00})} becomes (up to an alternating sign) the
$j$-body partition function for quantum particles with spectrum
$\epsilon _{k}$ and exclusion statistics parameter $g=2$, because of the
$+2$ shifts in the nested multiple sum {(\ref{00})}.

Indeed, consider now in place of ${{{\tilde{b}}}}_{p/q}(k)$ in
{(\ref{denote})} a general spectral function $s(k)$ and in place of
$a_{p,q}(2j)$ in {(\ref{00})} a general $n$-body partition
function\footnote{The notations $\mathbf{n}$ for the length --the
number of steps-- of closed lattice walks on the one hand, and $n$ for
the number of particles in statistical mechanics on the other hand,
should not be confused.} $Z(n)$ that rewrites as
%
\begin{eqnarray}
Z(n)
= \sum _{k_{1}=1}^{q-2n+2} \sum _{k_{2}=1}^{k_{1}} \cdots
\sum _{k_{n}=1}^{k_{n-1}}
s(k_{1}+2n-2)
s(k_{2}+2n-4) \ldots s(k_{n-1}+2)
s(k_{n})~.
\label{0}
\end{eqnarray}
Note that its relation to the Kreft coefficient in {(\ref{00})} is
$Z(j) = (-1)^{j-1} a_{p,q} (2j)$.

Again, due to the $+2$ shifts in the nested multiple sum {(\ref{0})}, the
arguments of the Boltzman factors $s(k)$ in the above expression differ
by at least 2; that is, no terms with particles in adjacent energy
levels $\epsilon _{k}$ and $\epsilon _{k+1}$ are admitted. This is
precisely the definition of the partition function for $n$ particles on
the line with energies $\epsilon _{k}$ obeying exclusion statistics of
order $g=2$. Explicitly, for the first few $Z(n)$ we have
\begin{eqnarray}
\nonumber
Z(1)
&=&\sum _{k_{1}=1}^{q} s(k_{1})\;,
\\
\nonumber
Z(2)
&=&\sum _{k_{1}=1}^{q-2}\sum _{k_{2}=1}^{k_{1}} s(k_{1}+2)s(k_{2})
\;,
\\
\nonumber
Z(3)
&=&\sum _{k_{1}=1}^{q-4} \sum _{k_{2}=1}^{k_{1}}\sum _{k_{3}=1}^{k
_{2}}s(k_{1}+4)s(k_{2}+2)s(k_{3})\;,
\\
\nonumber
Z(4)
&=&\sum _{k_{1}=1}^{q-6}\sum _{k_{2}=1}^{k_{1}} \sum _{k_{3}=1}^{k
_{2}}\sum _{k_{4}=1}^{k_{3}}s(k_{1}+6)s(k_{2}+4)s(k_{3}+2)s(k_{4})\;,
\\
\nonumber
{\mathrm{etc.}}
&&
\end{eqnarray}

If we set $s(k)=1$ -- i.e., we consider a degenerate quantum gas at
energy $0$ with $q$ single-particle quantum states -- we get
\begin{equation}
\nonumber
Z(n) = \frac{(q-n+1)!}{n!(q-2n+1)!} = \left . \binom{q-(g-1)(n-1)}{n}
\right |_{g=2}
\end{equation}
which is the number of ways to arrange $n$ particles in $q$
single-particle states on a line with exclusion statistics $g=2$. This
is the ``linear counting'' as originally proposed by Haldane
\cite{Haldane} where single-particle states are arranged on an open line
segment.

In an alternative ``periodic counting'' one starts instead with
single-particle states on a circle. $g=2$ exclusion statistics in this
case implies
%
\begin{eqnarray}
\nonumber
Z (n)
= \sum _{k_{1}=1}^{q-2n+2} \sum _{k_{2}=1}^{k_{1}} \cdots
\sum _{k_{n}=1}^{k_{n-1}}
&&{s}(k_{1}+2n-2)
{s}(k_{2}+2n-4) \cdots
{s}(k_{n}) 
\\
&&\bigg (1-\delta (k_{1}-(q-2n+2)) \; \delta (k_{n}-1)\bigg )
\label{opa}~~.
\end{eqnarray}
Again if we set ${s}(k)=1$ we obtain
\begin{equation}
\nonumber
Z(n) = \frac{q(q-n-1)!}{n!(q-2 n)!} = \left . \frac{q}{n}
\binom{q-(g-1)n-1}{n-1} \right |_{g=2}
\end{equation}
which is the number of ways to arrange $n$ particles in $q$ quantum
states on a circle with exclusion statistics $g=2$. This is the ``cyclic
counting'' that appeared in \cite{Poly} and is at work in the
thermodynamic limit of the LLL-anyon model \cite{Das}. Note that
for the specific spectral function {(\ref{denote})} of the square lattice
walk problem the two partition functions (open and periodic) are
identical, since $s(q)=\tilde{b}_{p/q}(q)=0$ and the extra terms in
{(\ref{0})} relative to {(\ref{opa})} vanish.

As is standard in many-body statistical mechanics, we introduce the
cluster coefficients $b(n)$ through the grand partition function
%
\begin{equation}
\log \left (\,\sum _{n=0}^{\infty }Z(n) z^{n} \right )=\sum _{n=1}^{
\infty } b(n) z^{n}
\label{grand}%
\end{equation}
with $z$ playing the role of fugacity; i.e.,
%
\begin{equation}
Z(n)=\sum _{{k_{j}\ge 0 \atop \sum _{j} jk_{j}=n}} \prod _{j=1}^{n}
\frac{1}{k_{j} !}b^{k_{j}}(j)~.
\label{0000}
\end{equation}
Explicitly, for the first few $n$,
\begin{eqnarray}
\nonumber
Z(1)
&=&b(1)
\\
\nonumber
Z(2)
&=&b(2)+\frac{1}{2!}b^{2}(1)
\\
\nonumber
Z(3)
&=&b(3)+b(1)b(2)+\frac{1}{3!}b^{3}(1)
\\
\nonumber
Z(4)
&=&b(4)+b(1)b(3)+\frac{1}{2!}b^{2}(2)+\frac{1}{2!}b^{2}(1)b(2)+
\frac{1}{4!}b^{4}(1)~.
\end{eqnarray}
The similarity to {(\ref{cluster})} and {(\ref{000})} is obvious -- as
already noticed the cluster coefficients in the Hofstadter case
are (up to alternating signs) the $q [q] a_{p,q} (2n)$'s,
$n\le {\mathbf{{n}}}/2$.

In the fermionic ($g=1$) and bosonic ($g=0$) cases the cluster
coefficients $b(n)$ are (up to sign) single-particle partition functions
with temperature parameter $n\beta $, that is, $\sum _{k=1}^{q} s^{n}
(k)$. In the $g=2$ exclusion case the corresponding expressions for the
cluster coefficients become
%
\begin{align}
\nonumber
b(1)
&=\sum _{k=1}^{q} s(k)\;,
\\
\nonumber
-b(2)
&=\frac{1}{2}\sum _{k=1}^{q} s^{2}(k)+\sum _{k=1}^{q} s(k+1)s(k)\;,
\\
\nonumber
b(3)
&=\frac{1}{3}\sum _{k}^{q} s^{3}(k)+\sum _{k=1}^{q} s^{2}(k+1)s(k)+
\sum _{k=1}^{q} s(k+1)s^{2}(k)
\\\nonumber
&+\sum _{k=1}^{q} s(k+2)s(k+1)s(k)\;,
\\
\nonumber
-b(4)
&= \frac{1}{4} \sum _{k=1}^{q} {s}^{4}(k)+ \sum _{k=1}^{q}
{s}^{3}(k+1) {s}(k)+ \sum _{k=1}^{q} {s}(k+1) {s}^{3}(k) +\frac{3}{2}
\sum _{k=1}^{q} {s}^{2}(k+1) {s}^{2}(k)
\\
\nonumber
& + 2 \sum _{k=1}^{q} {{s}}(k+2) {s}^{2}(k+1) {s}(k)+ \sum _{k=1}^{q}
{s}^{2}(k+2)
{s}(k+1) {{s}}(k) \\
&+ \sum _{k=1}^{q} {s}(k+2) {s}(k+1)
{s}^{2}(k)
 + \sum _{k=1}^{q} {s}(k+3) {s}(k+2) {s}(k+1) {s}(k)\label{tooth}
\end{align}
etc. The expressions (\ref{tooth}) generalize formula {(\ref{label})}
%
\begin{equation}
b(n)=(-1)^{n-1}
 \sum _{
{l_{1}, l_{2}, \ldots , l_{j} \atop { \mathrm{composition}}\;\mathrm{of}\;n}}
c(l_{1},l_{2},\ldots ,l_{j} )\sum _{k=1}^{q} {{{s}}}^{{l_{j}}}(k+j-1)
\ldots s^{l_{2}} (k+1) {s}^{l_{1}}(k)
\label{2}
\end{equation}
where the coefficients $c(l_{1},l_{2},\ldots ,l_{j})= c(l_{j},l_{j-1}
\dots , l_{1} )$ are given in {(\ref{ouf})}.
Formulae (\ref{tooth}) and (\ref{2}) hold both in the non-periodic (linear) counting
and the periodic counting (cyclic), provided that we put $s(k)=0$ for
$k>q$ in the non-periodic case and $s(k+q) = s(k)$ in the periodic case. Note also that in the periodic case for $q$ sufficiently small compared to $n$ additional corrections appear.\looseness=1

Formulae {(\ref{ouf})} and {(\ref{2})} were derived in \cite{nous} for
the ``building blocks'' of the algebraic area generating function
$[q]a_{p,q}(\mathbf{{n}})$ with the spectral function $s(k)$ being the
Hofstadter $\tilde{b}_{p/q}(k)$ in {(\ref{denote})}. Here we demonstrated
that they admit an interpretation in terms of $g=2$ exclusion
statistics. In fact, {(\ref{ouf})} and {(\ref{2})} provide the
\emph{microscopic} (that is, not the thermodynamic limit) $g=2$ cluster
coefficients for an arbitrary number $q$ of energy levels $\epsilon
_{k}$. Their form is intuitively clear: they ``correct'' the $n$-body
Boltzmann partition function $\bigl (\sum _{k=1}^{q} s (k)\bigr )^{n}/n!$
by subtracting terms forbidden by statistics. In the fermionic case
these are terms where particles occupy the same level~$k$, and they come
with coefficient $\pm 1/n$. In the $g=2$ case these are terms where the
particles occupy the same \emph{or neighboring} levels, as is clear by
the terms in $b(n)$ which are all in clusters of neighboring levels, and
they come with nontrivial combinatorial coefficients $\pm c(l_{1},
\dots ,l_{j} )$.

Setting as above $s(k)=1$ we obtain in the non-periodic case
\begin{equation}
b(n)= (-1)^{n-1} \; \frac{1}{2n} \left [ (q+2){\binom{2n }{n}}-2^{2n}
\right ]
\nonumber
\end{equation}
 and in the periodic case
\begin{equation}
b(n)=q (-1)^{n-1} \; \frac{1}{2n} {\binom{2n}{n}}
\nonumber
\end{equation}
i.e., (\ref{right}) up to sign. The large-$q$ limit of $b(n)$ gives the corresponding cluster
coefficients in the thermodynamic limit, $q$ playing the role of volume.
(Note that for the periodic case no limit is necessary.) The result
agrees with the known expressions obtained in \cite{Das,Poly} for the exclusion cluster coefficients
%
\begin{equation}
\label{g}
b(n) =q\frac{1}{n}\; \prod _{k=1}^{n-1}(1-g\, n /k)
\end{equation}
which, for $g=2$, reduces to $q(-1)^{n-1} {\binom{2n}{{n}}/ 2n}$.

\section{General $g$ microscopic cluster coefficients}%
\label{sec4}
We aim to generalize the statistics-lattice walks connection to a
general integer $g$. This goal involves two aspects: deriving the
expression for the microscopic cluster coefficients for exclusion
$g$ statistics, which were crucial in the statistics--Hofstadter model
connection, and then devising related random walk models on general
lattices (not necessarily square) that realize these statistics in terms
of algebraic area enumerations. In this section we address the first
aspect.

\subsection{$g=3 $}%
\label{sec4.1}
As a first step let us consider the case $g=3$ and its
cluster coefficients. Focusing on the periodic case (the non periodic
case can be treated along the same lines) the $n$-body partition
function becomes
%
\begin{eqnarray}
\nonumber
Z(n)
=
&& \sum _{k_{1}=1}^{q-3n+3} \sum _{k_{2}=1}^{k_{1}} \cdots
\sum _{k_{n}=1}^{k_{n-1}}
s(k_{1}+3n-3)
s(k_{2}+3n-6) \ldots s(k_{n-1}+3)
s(k_{n})
\\
&&\bigg (1-\delta (k_{1}-(q-3n+3))\delta (k_{n}-1)\bigg )\bigg (1-\delta
(k_{1}-(q-3n+2))\delta (k_{n}-1)\bigg )
\nonumber
\\
&&\bigg (1-\delta (k_{1}-(q-3n+3))\delta (k_{n}-2)\bigg ) \nonumber
\end{eqnarray}
(the awkward $\delta $-factors are there to eliminate terms that by
``umklapp'' would become nearest or next-to-nearest neighbors). The
cluster coefficients are defined again through the grand partition
function {(\ref{grand})}, the first few being
%
\begin{eqnarray}
\nonumber
b(1)=
&&\sum _{k=1}^{q} s(k)\;,
\\
\nonumber
-b(2)=
&&\frac{1}{2}\sum _{k=1}^{q} s^{2}(k)+\sum _{k=1}^{q} s(k+1)s(k)+
\sum _{k=1}^{q} s(k+2)s(k)\;,
\\
\nonumber
b(3)=
&&\frac{1}{3}\sum _{k}^{q} s^{3}(k)+\sum _{k=1}^{q} s^{2}(k+1) s(k)+
\sum _{k=1}^{q} s^{2}(k+2)s(k)
\\
\nonumber
&+&\sum _{k=1}^{q} s(k+1)s^{2}(k)+\sum _{k=1}^{q} s(k+2)s^{2}(k)
\\
\nonumber
&+&2\sum _{k=1}^{q} s(k+2)s(k+1)s(k)+\sum _{k=1}^{q} s(k+3)s(k+1)s(k)
\\
\label{oula}
&+&\sum _{k=1}^{q} s(k+3)s(k+2)s(k)+\sum _{k=1}^{q} s(k+4)s(k+2)s(k)
\end{eqnarray}
etc., where now jumps up to $2$ appear in the $k$-summation. The terms
in the cluster coefficients correct for terms in the Boltzmann partition
function where particles are in the same, nearest neighbor and
next-to-nearest neighbor states that are excluded by the $g=3$
statistics rule. Their general expression is still given by {(\ref{2})},
but now in the compositions of $n$ the occurrence of isolated $0$'s
(i.e., no two or more successive $0$'s and no $0$ in the first and last
entry) among the $l_{i}$'s is accepted.\footnote{There are
$3^{n-1}$ such compositions $l_{1}+l_{2}+\ldots +l_{j}=n$, $1\le j
\le 2n-1$ -- for example, for $n=3$ one has the $3^{2}=9$ compositions
$3=2+1=2+0+1=1+2=1+0+2=1+1+1=1+0+1+1=1+1+0+1=1+0+1+0+1$, with
$2+0+1$ leading to the term $\sum _{k=1}^{q} s^{2} (k+2) s(k)$,
$1+1+0+1$ to $\sum _{k=1}^{q} s(k+3) s(k+2) s(k)$, $1+0+1+0+1$ to
$\sum _{k=1}^{q} s(k+4) s(k+2) s(k)$, etc.}

In the $g=2$ case the coefficients $c(l_{1},l_{2},\ldots ,l_{j})$ 
in {(\ref{ouf})} can be rewritten as
\begin{equation}
\nonumber
c(l_{1},l_{2},\ldots ,l_{j})=
\frac{1}{l_{1}}\prod _{i=1}^{j-1}l_{i}\frac{ \bigg (
{l_{i}+l_{i+1} \atop l_{i},\; l_{i+1}}\bigg )}{l_{i}+l_{i+1}}~.
\end{equation}
In the $g=3$ case one finds the new set of coefficients $c(l_{1} ,
\dots , l_{j} )$'s to be
%
\begin{equation}
c(l_{1},l_{2},\ldots ,l_{j})= \frac{(l_{1}+l_{2}-1)!}{l_{1}!\;l_{2}!}
\prod _{i=1}^{j-2}\frac{l_{i}!\;l_{i+1}!}{(l_{i}+l_{i+1}-1)!}\frac{
\bigg ({l_{i}+l_{i+1}+l_{i+2} \atop l_{i},\; l_{i+1},\;l_{i+2}}
\bigg )}{l_{i}+l_{i+1}+l_{i+2}}
\label{nicenice}%
\;.
\end{equation}

Again, the $c(l_{1},l_{2},\ldots ,l_{j})$'s sum up to the $n$-th
thermodynamic cluster coefficient for particles with exclusion
statistics $g=3$ (up to an alternating sign)
\begin{equation}
{b(n) } = q(-1)^{n-1}
 \sum _{
{l_{1}, l_{2}, \ldots , l_{j} \atop {\mathrm{composition\;\mathrm{of}\;n}}\; \mathrm{with}\; \mathrm{isolated} \;0's}}
 c(l_{1},l_{2},\ldots ,l_{j})=
q\frac{1}{n}\; \prod _{k=1}
^{n-1}{(1-g\, n/k)}
\nonumber
\end{equation}
which for $g=3$ reduces to
%
\begin{equation}
\label{33}
b(n)=q(-1)^{n-1} \frac{1}{3n}{{\binom{3n}{n}}}\;.
\end{equation}

\subsection{Generalization to higher $g$}%
\label{sec4.2}
The above results can be generalized to higher integer exclusion
statistics $g$. The interesting and novel quantities are the microscopic
cluster coefficients $b(n)$. These are given by {(\ref{2})}, as for
$g=2$ and $g=3$, but now with a generalized definition of compositions
and new expressions for the $c(l_{1},l_{2},\ldots ,l_{j})$'s.

Let us define a $g$-composition of a positive integer $n$ as the ordered
set of any number of positive or zero integers $l_{1},l_{2},\ldots ,l
_{j}$ such that $l_{1} + \dots + l_{j} = n$ and where at most $g-2$
consecutive numbers can be zero, therefore extending the compositions
admitted in the $g=2$ and $g=3$ cases ($2$-compositions are the standard
compositions with nonzero entries, while $1$-compositions would be
defined as the unique integer $n$). There are $g^{n-1}$ such
compositions in total.

The $g$-cluster coefficients $b (n)$, then, are given by the general
formula
%
\begin{equation}
b (n)=(-1)^{n-1}
 \sum _{
{l_{1}, l_{2}, \ldots , l_{j} \atop g\mathrm{-composition}\;\mathrm{of}\;n}}
 c (l_{1},l_{2},\ldots ,l_{j} )\sum _{k=1}^{q} {{{s}}}^{{l_{j}}}(k+j-1)
\ldots s^{l_{2}} (k+1) {s}^{l_{1}}(k)
\label{genc}\nonumber
\end{equation}
with the coefficients $c (l_{1},l_{2},\ldots ,l_{j})$, $1\le j
\le (g-1)(n-1)+1$, given by
%
\begin{eqnarray}
{c (l_{1},l_{2},\ldots ,l_{j})}
&=& {{\frac{(l_{1}+\dots +l_{g-1}-1)!}{l
_{1}! \cdots l_{g-1}!}~
\prod _{i=1}^{j-g+1}
\binom{l_{i}+\dots +l_{i+g-1}-1 }{l_{i+g-1}}}}
\nonumber
\\
&=& {{\frac{\prod _{i=1}^{j-g+1} (l_{i} + \dots + l_{i+g-1} -1)!
}{
\prod _{i=1}^{j-g} (l_{i+1} + \dots +l_{i+g-1} -1 )! } }
\prod _{i=1}
^{j} \frac{1}{l_{i}!} } ~~.\label{glong}
\end{eqnarray}
The second expression in (\ref{glong}) makes explicit the fact that the coefficients are
symmetric under reversal of order $c (l_{1},l_{2},\ldots ,l_{j}) =
c (l_{j}, \dots ,l_{2},l_{1} )$ since each product in it is
manifestly invariant under this reversal. It is the generalization to
any higher $g$ of {(\ref{ouf})} for $g=2$ and {(\ref{nicenice})} for $g=3$.
Again, as for $g=2,3$ in {(\ref{g},\ref{33})}, in the degenerate case the sum of these
coefficients gives (up to an alternating sign) the thermodynamic cluster
coefficient \cite{Das,Poly} of $n$ particles with exclusion
statistics $g$, that is,
%
\begin{equation}
b (n) = q(-1)^{n-1} \frac{1}{gn} {\binom{gn }{n}} \nonumber
\end{equation}

\section{Matrix formulation of exclusion statistics}%
\label{sec5}

We now turn to the connection between exclusion statistics and various
lattice models with appropriate random walk algebraic area enumeration.
Since the connection between the Hoftstadter model and random walks on
the square lattice for $g=2$ exclusion statistics was established from
the Hofstadter matrix {(\ref{matrix})} and its secular determinant, we
focus on the structure of this and related matrices.

\subsection{$g=2$: Hofstadter-like matrix}%
\label{sec5.1}
The Hofstadter matrix {(\ref{matrix})} has a form not very convenient for
computing its determinant. In particular, the spectral function
${\tilde{b}}_{p,q} (k)$ does not show up explicitly in the elements of
the Hamiltonian {\eqref{Hof}}, with $u$ and $v$ given in {\eqref{umat}}, {\eqref{vmat}} ($k_{x}=k_{y}=0$
understood).

In \cite{Kreft} an alternative form of the Hamiltonian was used,
which consisted of a transformation of the matrices $u$ and $v$ that
preserves their algebra
\begin{equation}
\nonumber
u \to - u\; v ~,
~~~
v \to v~.
\end{equation}
This transformation is essentially unitary, as guaranteed
by the facts that it preserves the commutation relation $v \, u =
\mathrm{Q}\, u \, v$ and that the given matrix realization of $u$ and
$v$ is an irreducible representation of this relation. The Casimir
$u^{q}$, however, is mapped to
\begin{equation}
\nonumber
u^{q} \to (-u v)^{q} = (-1)^{q} \, \mathrm{Q}^{q(q-1)/2} = -
(-1)^{(p+1)(q-1)} = -1
\end{equation}
(the last equality follows from the fact that at least one of $p$,
$q$ must be odd, since they are co-prime, and therefore $(p+1)(q-1)$ is
always even). The transformation $u \to -uv$ is thus unitary up to a
phase. This, and any similar change of the Casimirs, has no effect on
the counting of closed walks and only affects walks with at least
$q$ steps in the horizontal direction that are not closed but are still
counted by the Hofstadter Hamiltonian, that is, spurious ``umklapp''
terms.

Transformations preserving the commutation relation are, in general,
area-preserving lattice automorphisms that deform individual random
walks but leave their algebraic area, and the number of walks
corresponding to a given area, invariant.

Under the above transformation the Hofstadter Hamiltonian {(\ref{Hof})} rewrites
as
%
\begin{equation}
H'=-u\;v -v^{-1}\;u^{-1}+v+v^{-1} ~.
\label{newHof}%
\end{equation}
In this new representation the matrix
%
\begin{equation}
1-z H'
=
{\fontsize{8.8}{9}
\selectfont
\begin{pmatrix}
1& -(1-\mathrm{Q})z & 0 & \cdots & 0 & -(1-{\mathrm{Q}^{-q}})z
\\
-(1-{\mathrm{Q}}^{-1})z & 1 & -(1-\mathrm{Q}^{2})z & \cdots & 0 & 0
\\
0 & -(1-{\mathrm{Q}^{-2}})z & 1 &\cdots & 0 & 0
\\
\vdots & \vdots & \vdots & \ddots & \vdots & \vdots
\\
0 & 0 & 0 & \cdots & 1 & -(1-{\mathrm{Q}^{q-1}})z
\\
-(1-\mathrm{Q}^{q})z & 0 & 0 & \cdots & -(1-{\mathrm{Q}^{1-q}})z& 1
\\
\end{pmatrix}}
\nonumber
\end{equation}
possesses two important properties:
\begin{enumerate}
\item It has unit diagonal elements.
\item It is tridiagonal, since $\mathrm{Q}^{q}=1$ and thus the corner
elements $1-\mathrm{Q}^{-q}$ and $1-\mathrm{Q}^{q}$ vanish.%
\end{enumerate}
The latter property facilitates the calculation of the determinant,
yielding
\begin{equation}
\det (1-z H')=-\sum _{j=0}^{\lfloor q/2 \rfloor }a_{p, q}(2j)z^{2j}
\nonumber
\end{equation}
where the umklapp term $-4z^{q}$ in {(\ref{matrix})} is now
absent. Also, the Hofstadter spectral function ${\tilde{b}_{p/q}}(k)$
is actually recovered as the product of the off-diagonal elements
\begin{equation}
{\tilde{b}_{p/q}}(k)= (1-\mathrm{Q}^{k} )(1-\mathrm{Q}^{-k} ) = 4 \sin
^{2}(\pi k p/q)\;.
\nonumber
\end{equation}

Motivated by these observations, let us define in the $g=2$ case a
general Hamiltonian $H_{2}$ of the above cyclic type but with general
off-diagonal elements
\begin{equation}
\nonumber
(H_{2})_{ij}= \mathbf{f}(i) \, \delta _{i+1,j} + \mathbf{g}(j) \,
\delta _{j+1,i} ~,
~~
{\mathrm{with}}
~~
\delta _{i,i+q}=1 ~~{\mathrm{understood}}
\end{equation}
or, explicitly,
%
\begin{equation}
H_{2} =
\begin{pmatrix}
0& \mathbf{f}(1) & 0 & \cdots & 0 & \mathbf{g}(q)
\\
\mathbf{g}(1) & 0 & \mathbf{f}(2) & \cdots & 0 & 0
\\
0 & \mathbf{g}(2) & 0 &\cdots & 0 & 0
\\
\vdots & \vdots & \vdots & \ddots & \vdots & \vdots
\\
0 & 0 & 0 & \cdots & 0 & \mathbf{f}(q-1)
\\
\mathbf{f}(q) & 0 & 0 & \cdots & \mathbf{g}(q-1) & 0
\\
\end{pmatrix}
\label{matrix2t}~~. \nonumber%
\end{equation}
Hermiticity of $H_{2}$ would require $\mathbf{f}(k) = \mathbf{g}(k)^{*}$
but we allow in general for non-Hermitian Hamiltonians, so $
\mathbf{f}(k)$ and $\mathbf{g}(k)$ are arbitrary complex numbers. Note,
further, that $\mathbf{f}(q)=0$ or $\mathbf{g}(q)=0$ is not required.

Our basic observation is that the secular determinant of the matrix
$1-z H_{2}$ gives the grand partition function for particles of
exclusion statistics $g=2$ with spectral function $s_{2} (k)$ and
fugacity $z_{2}$ given by
\begin{equation}
\nonumber
s_{2} (k) = \mathbf{g}(k) \mathbf{f}(k)  ~,
~~~
z_{2} = - z^{2}
\end{equation}
up to a spurious umklapp term at order $z^{q}$. Specifically, we get
%
\begin{align}
\nonumber
\det (1-z H_{2} ) = \, 1
&- z^{2} \sum _{k_{1}=1}^{q} s_{2} (k_{1}) + z
^{4}\, \sum _{k_{1}=1}^{q-2}\, \sum _{k_{2}=1}^{k_{1}} s_{2} (k_{1}+2) s
_{2} (k_{2})\\\nonumber
&\times\bigg (1-\delta (k_{1}-(q-2))\delta (k_{2}-1)\bigg )
\\
\nonumber
&-z^{6} \, \sum _{k_{1}=1}^{q-4} \sum _{k_{2}=1}^{k_{1}} \, \sum _{k_{3}=1}
^{k_{2} }s_{2} (k_{1}+4) s_{2} (k_{2}+2) s_{2} (k_{3})
\\\nonumber
&\times
 \bigg (1-\delta
(k_{1}-(q-4))\delta (k_{3}-1)\bigg )
\\
\nonumber
& + \dots
\\
\label{H2det}
&+ (-z)^{q} \left (\,\prod _{k=1}^{q} {\mathbf{f}}(k) + \prod _{k=1}^{q}
{\mathbf{g}}(k) \right )
\end{align}
as in {(\ref{opa})}, or, in another writing,
%
\begin{align}
\nonumber
\det (1-z H_{2} ) = \, 1
&- z^{2} \sum _{k_{1}=1}^{q} s_{2} (k_{1}) + z
^{4}\, \sum _{k_{1}=1}^{q-2}\, \sum _{k_{2}\ge {\mathrm{max}}(1,k_{1}-q+4)}
^{k_{1}} s_{2} (k_{1}+2) s_{2} (k_{2})
\\
\nonumber
&-z^{6} \, \sum _{k_{1}=1}^{q-4} \sum _{k_{2}=1}^{k_{1}} \,
\sum _{k_{3}\ge {\mathrm{max}}(1,k_{1}-q+6)}^{k_{2} }s_{2} (k_{1}+4) s
_{2} (k_{2}+2) s_{2} (k_{3})
\\
\nonumber
& + \dots
\\
\label{H2detbis}
&+ (-z)^{q} \left (\,\prod _{k=1}^{q} {\mathbf{f}}(k) + \prod _{k=1}^{q}
{\mathbf{g}}(k) \right ) ~~.\nonumber
\end{align}
The term of order $(-z^{2})^{n}$ is the partition function of $n$
particles on $q$ energy levels $\epsilon _{k}$ determined by
$e^{-\beta \epsilon _{k}} = s_{2} (k)$ with $g=2$ exclusion on the
circle. The last term is the spurious umklapp one, the only term in
which $\mathbf{f}(k)$ and $\mathbf{g}(k)$ do not appear in the
combination $\mathbf{f}(k) \mathbf{g}(k)=s_{2} (k)$.

Notice that if at least one $\mathbf{f}(k)$ and one $\mathbf{g}(k)$
vanish the umklapp term disappears. Further, if at least one
$s_{2} (k)$ vanishes, which can be chosen to be $s_{2} (q)$ by a cyclic
renaming of states, then the cyclic constraint in the summation becomes
irrelevant and the result coincides with the linear exclusion counting.
Both the above conditions hold for the Hofstadter spectral function
$s_{2} (k) = {\tilde{b}_{p/q}}(k)$ given in {(\ref{denote})}.

\subsection{Generalization to higher $g$}%
\label{sec5.2}
The construction of the previous section can be generalized to matrices
that reproduce higher $g$ exclusion statistics: consider the cyclic
$H_{g}$ matrix
%
\begin{equation}
(H_{g})_{ij}= \mathbf{f}(i) \, \delta _{i+1,j} + \mathbf{g}(j) \,
\delta _{j+g-1,i} ~,
~~
{\mathrm{with}}
~~
\delta _{i,i+q}=1
\label{Hg}%
\end{equation}
with two nonzero off-diagonals (plus their `tails' as they wrap around
the length of the matrix), one immediately above the main diagonal and
the other $g$ positions below it. The matrix {(\ref{Hg})} obviously cannot
be Hermitian, but it is acceptable for generating lattice random walks,
as we shall see.

The basic fact is that the secular determinant of the matrix
$1- z H_{g}$ reproduces, up to umklapp terms, the grand partition
function of exclusion-$g$ particles with spectral function $s_{g} (k)$
and fugacity $z_{g}$ given by
\begin{equation}
\nonumber
s_{g} (k) = \mathbf{g}(k)\, \mathbf{f}(k) \mathbf{f}(k+1) \cdots
{\mathbf{f}}(k+g-2) ~,
~~~
z_{g} = - z^{g}
\end{equation}
that is,
\begin{align}
\nonumber
\det (1-z H_{g} ) = \, 1
&- z^{g} \sum _{k_{1}=1}^{q} s_{g} (k_{1}) + z
^{2g}\, \sum _{k_{1}=1}^{q-g}\,
\sum _{k_{2}\ge {\mathrm{max}}(1, k_{1}-q+2g)}^{k_{1}} s_{g} (k_{1}+g)
s_{g} (k_{2})
\\
\nonumber
&-z^{3g} \, \sum _{k_{1}=1}^{q-2g} \sum _{k_{2}=1}^{k_{1}} \,
\sum _{k_{3}\ge {\mathrm{max}}(1,k_{1}-q+3g)}^{k_{2} }s_{g} (k_{1}+4) s
_{g} (k_{2}+2) s_{g} (k_{3})
\\
&+ \dots
\nonumber
\\
&+ (-z)^{n_{c}} ~{\mathrm{{(umklapp ~ terms)}}}~.
\label{detzHg}
\end{align}
It is of no interest to give an explicit expression for the umklapp
terms. Such terms appear at power $z^{n_c}$ with
%
\begin{equation}
n_{c} = g \Bigl \lceil \raisebox{0.03cm}{$\displaystyle\frac{q}{g-1}$} \Bigr \rceil - q
\label{nc}%
\end{equation}
where $\lceil ~\rceil$ denotes the `ceiling' function. So the largest number of particles in the grand partition function
reproduced faithfully by the determinant is
\begin{equation}
\nonumber
\Bigl \lceil \raisebox{0.03cm}{$\displaystyle\frac{q }{g-1}$} \Bigr \rceil-\Bigl \lfloor \raisebox{0.05cm}{$\displaystyle\frac{q}{g}$} \Bigr \rfloor -1  ~.
\nonumber
\end{equation}
Note that for the $g=2$ Hofstadter-like case $n_{c}=q$, as explicitly
displayed in {(\ref{H2det})}, while the largest number of particles
$\lceil q/2\rceil-1$ is almost the same as the maximum allowed by exclusion statistics,
namely $\lfloor q/2 \rfloor$, while for $g\ge 3$ umklapp terms appear
before the maximum number of particles, $\lfloor q/g\rfloor $, is
reached.

We also point out that if $g-1$ successive values of $s_{g} (k)$ vanish,
say $s_{g} (1) = \dots = s_{g} (g-1) = 0$, the cyclic and linear
exclusion (i.e. periodic versus non periodic) statistics countings
coincide. If, moreover, the stronger condition
\begin{equation}
\nonumber
{\mathbf{g}}(1) = \dots = \mathbf{g}(g-1) = 0 ~,
~~~
{\mathbf{f}}(g-1) =0
\end{equation}
holds, the umklapp terms actually vanish. (Note that $\mathbf{f}(g-1)=0$
is the only $\mathbf{f}(k)$ appearing in the definition of $s_{g} (1)
, \dots , s_{g} (g-1)$ and in no other $s_{g} (k)$.) In that case the
secular determinant has only two nonzero off-diagonals (the terms in the
wraparound `tails' vanish) and gives the full exact exclusion-$g$ grand
partition function.

Finally, we point out that the secular determinant $\det (1-z H_{g} )$
admits two distinct statistical mechanical interpretations:
\begin{enumerate}
\item As a \emph{fermionic} grand partition function with fugacity
parameter $z_{1} = -z$ and particles occupying energy levels
$\varepsilon _{k}$ given by $e^{-\beta \varepsilon _{k}} = E_{k}$, where
$E_{k}$ are the eigenvalues of $H_{g}$.

\item As an \emph{exclusion statistics $g$} grand partition function with
fugacity parameter $z_{g} = -z^{g}$ and levels $e^{-\beta \epsilon
_{k}} = s_{g} (k)$, as in (\ref{detzHg}).
\end{enumerate}

The fact that the same system admits these two distinct interpretations
is a kind of generalized ``bosonization'' (really a
``superfermionization'') that maps an exclusion-1 (fermion) system to an
exclusion-$g$ system. In that trade-off the spectrum of the exclusion-$g$ 
system is simple and known while the spectrum of the fermionic
system is nontrivial (e.g., the Hofstadter Hamiltonian spectrum in the
$g=2$ case).

\section{Random walks corresponding to higher exclusion statistics}%
\label{sec6}

We proceed, now, to identify random walks that realize exclusion
statistics $g$ through their algebraic area enumeration generating
function, as was the case for square lattice walks with $g=2$ statistics.

\subsection{General construction}%
\label{sec6.1}

Let us rewrite the Hamiltonian {(\ref{Hg})} corresponding to exclusion
statistics $g$ in terms of $u$ and $v$ in {(\ref{umat})} and {(\ref{vmat})}
($k_{x}=k_{y}=0$ understood)
%
\begin{equation}
H_{g} = F(u) \; v + v^{1-g} \; G(u)
\label{Huv}
\end{equation}
where $F(u)$ and $G(u)$ are scalar functions of $u$ such that
\begin{equation}
\nonumber
F(\mathrm{Q}^{k} ) = \mathbf{f}(k)~,
~~~
G(\mathrm{Q}^{k} ) = \mathbf{g}(k) ~.
\end{equation}
This puts us in the context of random walks on the square lattice, assuming
$u$ and $v$ represent unit hops in the horizontal and vertical
directions, but with allowed steps as defined in the above Hamiltonian
upon expanding $F(u)$ and $G(u)$ as a power series in $u$. E.g.,
choosing $F(u)= 1-u$, $G(u)=1-u^{-1}$ and $g=2$ we recover the
Hofstadter Hamiltonian {(\ref{newHof})}.

We also note that the results stated in section~\ref{sec5.2} can be
systematically obtained in the above representation by writing
%
\begin{equation}
\det (1-z H_{g} ) = e^{\mathrm{Tr}\:
 \ln (1-z H_{g} )} = \exp \left (-\sum _{n=1}^{\infty }\frac{z
^{n} }{n} {\mathrm{Tr}}\:H_{g}^{n}\right ) \nonumber
\end{equation}
and taking into account that $v^{q} =1$ and that, for any scalar
function $X(u)$,
\begin{equation}
{\mathrm{Tr}}\:\left ( X(u) v^{n} \right ) = \sum _{k=-\infty }^{
\infty }\delta _{n-k q} ~{\mathrm{Tr}}\:X(u) ~.
\nonumber
\end{equation}
 Terms with $n=0$ reproduce the exclusion-$g$ grand partition function
and start at power $z^{g}$, while terms with $n=\pm q, \dots $ produce
the umklapp effects and start at $z^{n_{c}}$ as in {(\ref{nc})}.

The identification of lattice random walks that realize fractional
statistics $g$ relies not only upon choosing specific spectral functions
$F(u)$ and $G(u)$ in the above Hamiltonian, but also upon choosing
specific realizations of the matrices $u$ and $v$ in terms of lattice
hopping operators. The obvious choice is to interpret $u$ and $v$ as the
horizontal and vertical hopping operators. However, other choices may
yield a different lattice and a different set of walks. Although all
these choices are related by area-preserving lattice mappings, and thus
give the same area counting, some interpretations may be more symmetric,
more physically motivated and more useful. We see this pattern in the
Hofstadter case where {(\ref{newHof})} corresponds to vertical hops and
hops in the $45^{\mathrm{o}}$ diagonal, while the original realization
{(\ref{Hof})} gives the standard random walks on the square lattice.

\subsection{$g=3$ and chiral walks on the triangular lattice}%
\label{sec6.2}

Let us illustrate the above considerations in the case $g=3$ with a
choice of spectral functions and realizations that will map to chiral
walks on a triangular lattice.

The starting Hamiltonian is
%
\begin{equation}
\label{triangular}
H_{t} = U + V + \mathrm{Q}^{1+a} \; U^{-1} V^{-1}~,
\end{equation}
with $a$ an arbitrary real number. The unitary matrices $U$, $V$ satisfy
%
\begin{equation}
V\, U = \mathrm{Q}^{2} \, U \, V
\label{Qsquare}%
\end{equation}
where the change from $\mathrm{Q}$ to $\mathrm{Q}^{2}$ is for later
convenience. Note that any other phase factors introduced in the above
Hamiltonian could be absorbed in appropriate redefinitions of the phases
of $U$ and $V$, so $a$ and $\mathrm{Q}$ are the only relevant
parameters in the problem.

This Hamiltonian is clearly not (yet) of the general form {(\ref{Huv})}.
Before bringing it to that form let us give it a lattice interpretation
as acting on a particle hopping on the vertices of a triangular lattice.
$U$, $V$ and $W=\mathrm{Q}^{1+a} \, U^{-1} V^{-1}$ correspond to motion
in directions with angles $0$, $2\pi /3$ and $4\pi /3$, respectively,
with respect to the horizontal axis. Therefore, $W V U= V U W= U W V =
\mathrm{Q}^{1+a}$ corresponds to a closed counterclockwise walk along
the three sides of an elementary \emph{up-vertex} triangular cell, and
it assigns to this cell an area $1+a$. Conversely, $V W U = W U V = U
V W =\mathrm{Q}^{-1+a} = (\mathrm{Q}^{1-a})^{-1}$ represents a closed
\emph{clockwise} walk around an elementary \emph{down-vertex} triangular
cell, and it assigns to this cell an area $1-a$ (see {Fig.~\ref{fig1}}). 
\begin{figure}
\hskip 4.5cm
\includegraphics[scale=.25]{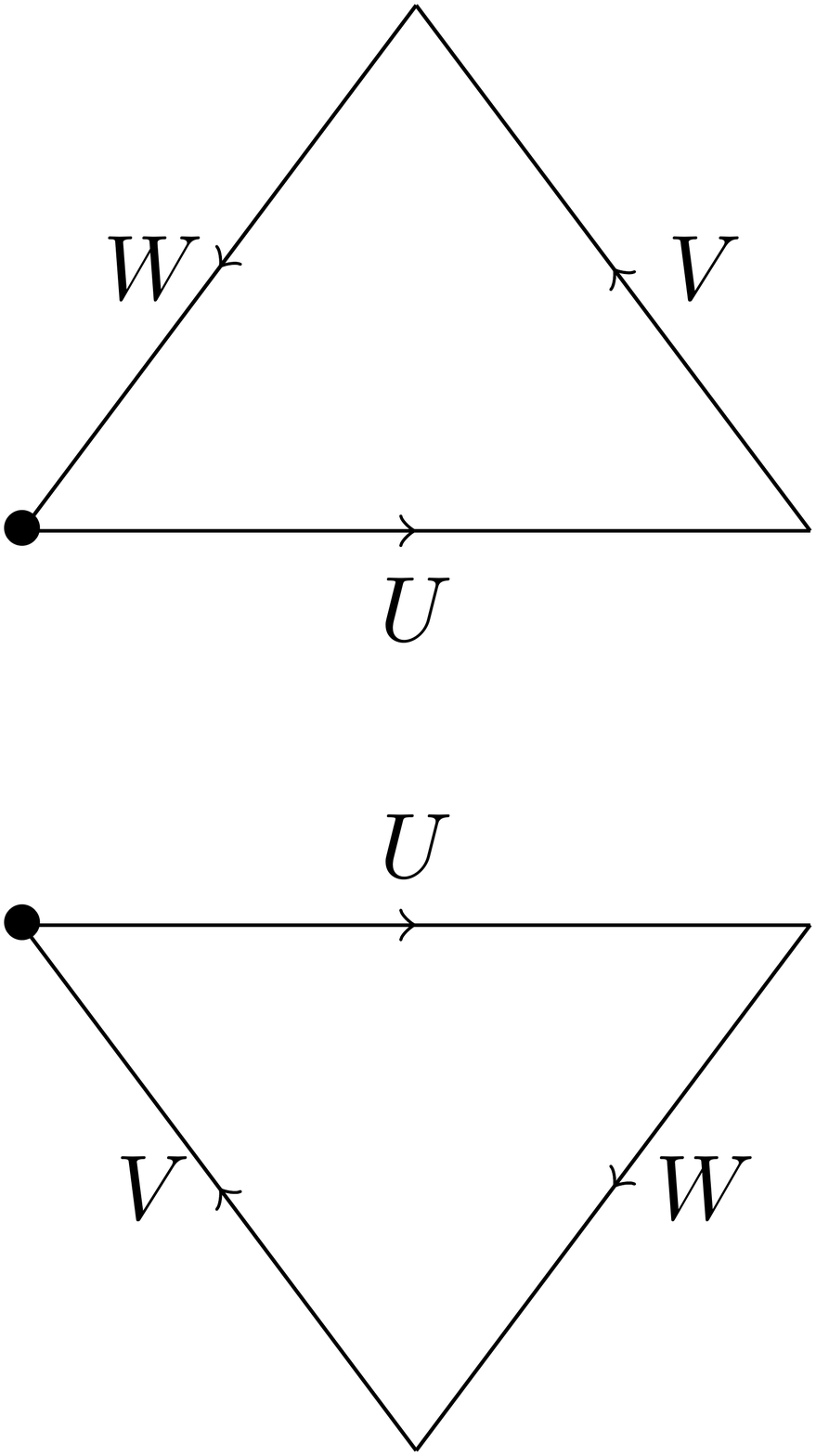}\label{fig1}
\caption{Walks going around \emph{up-vertex} and \emph{down-vertex} triangular
cells starting from the black bullet lattice site.}\label{fig1}
\end{figure}
So, in
general, up-vertex and down-vertex triangular cells can have
different areas, and $a$ is the half-difference of the area of the
up-vertex and down-vertex triangular cells. Alternately, we can
consider $\mathrm{Q}_{u} = \mathrm{Q}^{1+a}$ and $\mathrm{Q}_{d} =
\mathrm{Q}^{1-a}$ as counting parameters for the algebraic number of
up-vertex and down-vertex triangular cells enclosed by the walks. If we wish
to assign up-vertex and down-vertex triangular cells the same area,
so that they be counted on an equal footing, we must take $a=0$.

Powers of the Hamiltonian {(\ref{triangular})} represent random
\emph{chiral} walks on the triangular lattice: from each lattice vertex
the walk can proceed in only three of the possible six directions (see
{Fig.~\ref{fig2}}).
%
\begin{figure}
\vskip -3cm
\hskip 4.5cm\includegraphics[scale=.35]{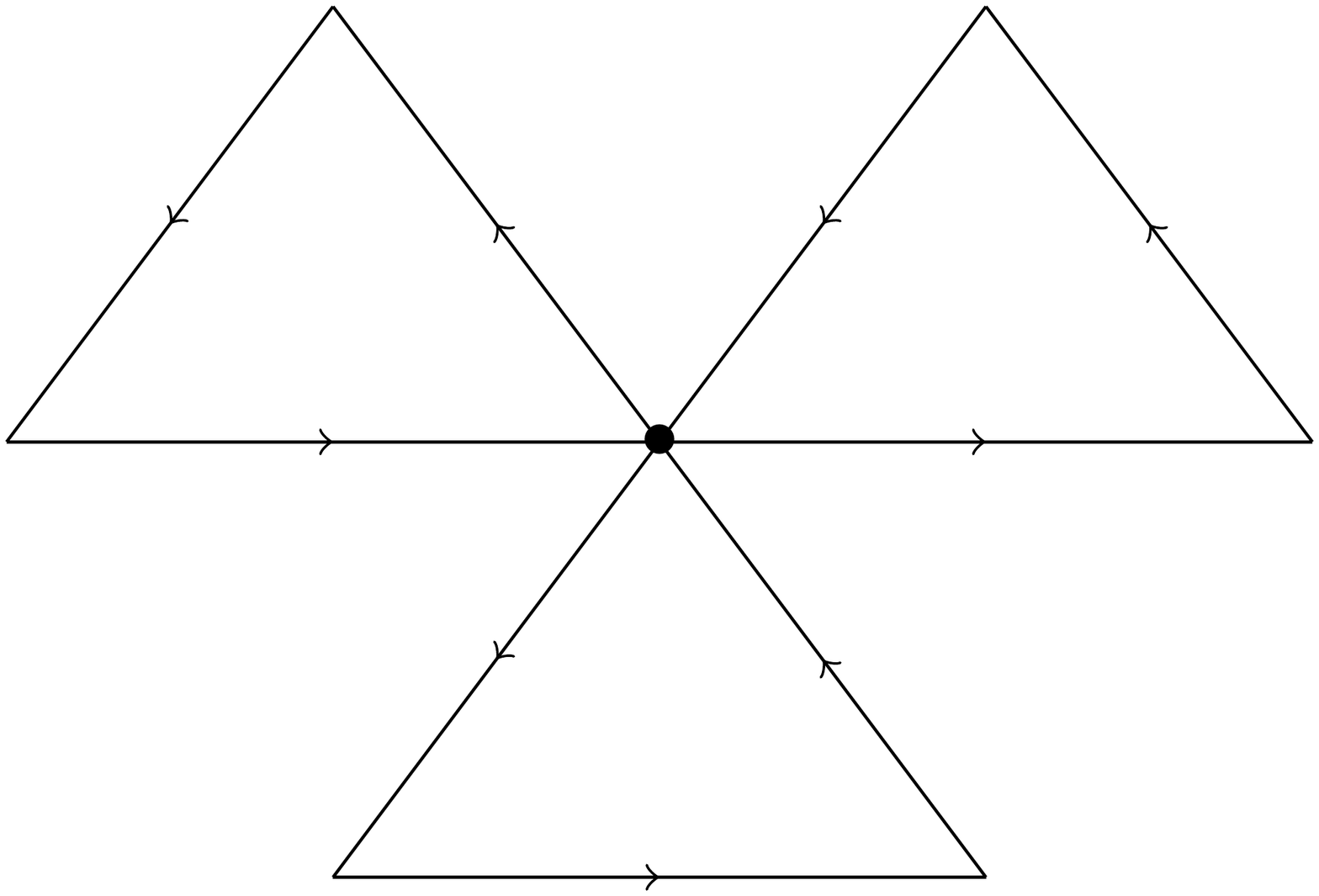}\label{fig2}
\vskip -2cm
\caption{Three of the 6 possible chiral walks starting from the same black
bullet lattice site. Only the 3 outgoing arrows represent possible motions from
the original site.}\label{fig2}
\end{figure}
\begin{figure}
\hskip 4.5cm
\includegraphics[scale=.35]{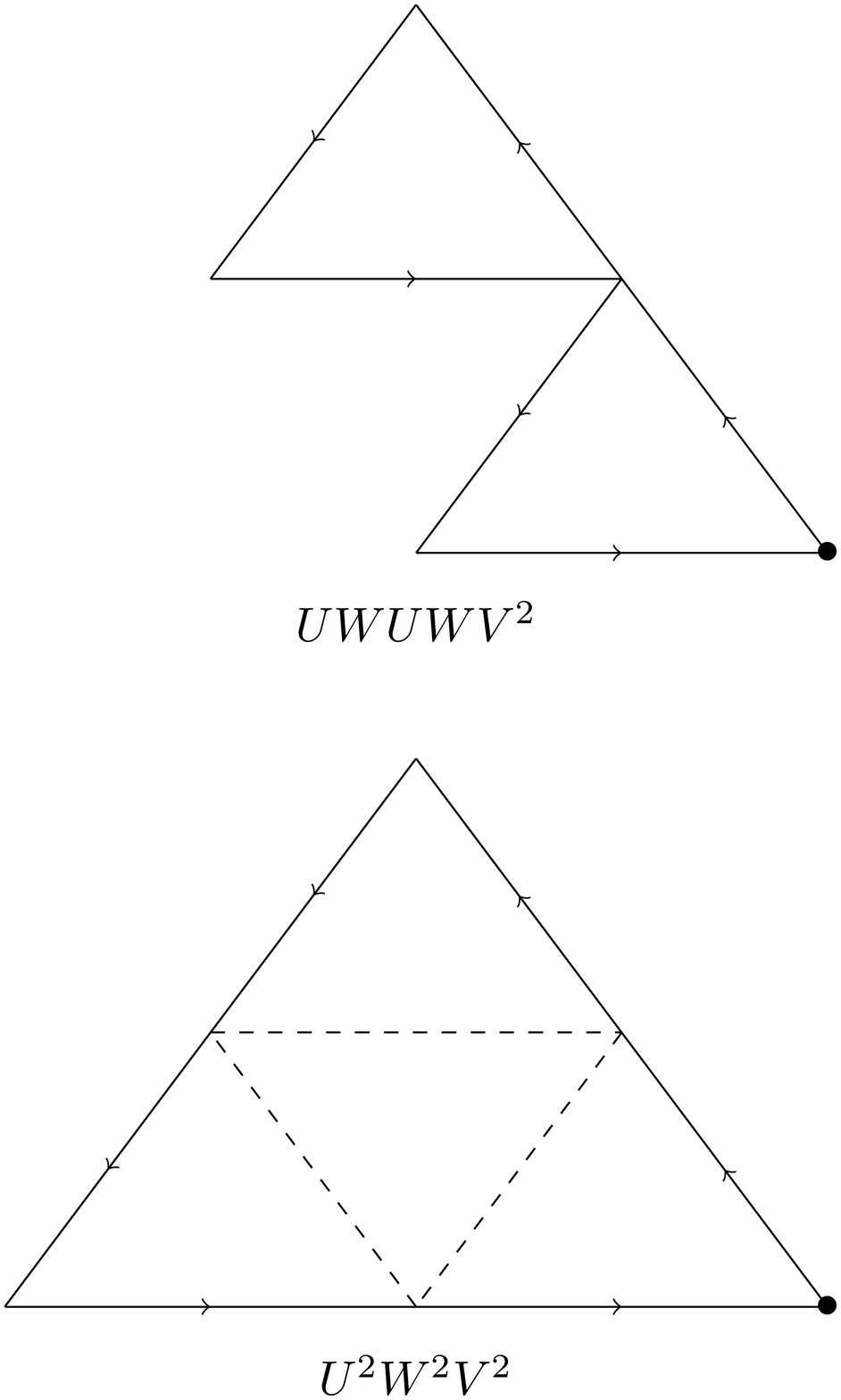}\label{fig3}
\caption{$UWUWV^{2}$ and $U^{2} W^{2} V^{2}$ walks.}\label{fig3}
\end{figure}
E.g., $U W U W V^{2} = \mathrm{Q}^{2(1+a)}$ is a closed walk enclosing
two up-vertex triangular cells, each in the counterclockwise sense,
while $U^{2} W^{2} V^{2} =\mathrm{Q}^{4+2a}= \mathrm{Q}^{3(1+a)}
\mathrm{Q}^{1-a}$ is a closed walk enclosing 3 up-vertex and one
down-vertex triangular cells, all in the counterclockwise sense (see
{Fig.~\ref{fig3}}).

Clearly only walks with a total number of steps $\mathbf{{n}}$ a
multiple of 3 can be closed. As in square lattice walks, $\mathrm{Tr}\:H_{t}^{\mathbf{{n}}}$ is the generating function of closed
walks of length $\mathbf{{n}}=3n$ weighted by the exponential of their
algebraic area, up to umklapp effects.

To bring the triangular lattice Hamiltonian {(\ref{triangular})} to the
standard exclusion form {(\ref{Huv})} we choose the representation
\begin{equation}
\nonumber
U = -i \, u \, v ~,
~~~
V = i\, u^{-1} \, v
\end{equation}
with $u$, $v$ the standard matrices {(\ref{umat})} and {(\ref{vmat})} ($k
_{x}=k_{y}=0$ understood), which reproduces the commutation relation
{(\ref{Qsquare})}. (It also makes the Casimirs $U^{q} = i^{q}$ and
$V^{q} = (-i)^{q}$ but, as stated before, this only affects umklapp
terms.) $H_{t}$ in {(\ref{triangular})} then becomes
\begin{equation}
\nonumber
H_{3} = i(- u+ u^{-1})\, v + \mathrm{Q}^{a} \, v^{-2}
\end{equation}
which is of the exclusion form {(\ref{Huv})} with
\begin{equation}
\nonumber
F = i(- u + u^{-1}) ~,
~~~
G = \mathrm{Q}^{a} ~,
~~~
g=3~.
\end{equation}
We obtain the spectral parameters
\begin{equation}
\nonumber
{\mathbf{f}}(k)= i\left(-\mathrm{Q}^k +\mathrm{Q}^{-k}\right) = 2 \sin ({2 \pi k p/q}) ~,
~~~
{\mathbf{g}}(k)= e^{2i\pi a p/q}
\end{equation}
and the spectral function
%
\begin{equation}
s_{3} (k) = \mathbf{g}(k) \, \mathbf{f}(k) \, \mathbf{f}(k+1) = 4 e
^{2i\pi a p /q} \sin ({2 \pi k p/q})
\sin
({2 \pi (k+1)p/q})
\label{OK}
\end{equation}
while the corresponding secular matrix assumes the form
%
\begin{equation}
1-z H_{3}=
{\fontsize{7.5}{7}\selectfont
\begin{pmatrix}
1& i(\mathrm{Q}-\frac{1}{\mathrm{Q}})z & 0 & \cdots & 0 &-\mathrm{Q}
^{a} z &0
\\
0 & 1 & i(\mathrm{Q}^{2}-\frac{1}{\mathrm{Q}^{2}})z &\cdots & 0&0& -
\mathrm{Q}^{a} z
\\
-\mathrm{Q}^{a} z & 0 & 1 &\cdots &0 & 0&0
\\
\vdots &\vdots & \vdots &\ddots &\vdots &\vdots & \vdots
\\
0 &0 & 0 & \cdots &1 &
i(\mathrm{Q}^{q-2}-\frac{1}{\mathrm{Q}^{q-2}})z&0
\\
0 & 0 & 0 &\cdots & 0&1 &i(\mathrm{Q}^{q-1}-
\frac{1}{\mathrm{Q}^{q-1}})z
\\
i(\mathrm{Q}^{q}-\frac{1}{\mathrm{Q}^{q}})z & 0 & 0 & \cdots &-
\mathrm{Q}^{a} z&0 & 1
\\
\end{pmatrix}}~~.
\nonumber
\end{equation}
Note that, since $s_{3} (q-1) = s_{3} (q) =0$, the periodic and linear
exclusion statistics countings coincide, as in the square lattice case.

It follows that on a triangular lattice the algebraic area enumeration
generating function for closed chiral random walks is obtained as the
grand partition function of exclusion-3 particles with the spectral
function $s_{3} (k)$ given in {(\ref{OK})}. (The secular determinant will
also produce umklapp terms, but these correspond to walks that are not
closed but are counted as such and can simply be ignored.)

For closed walks of length $\mathbf{{n}}=3n$, $s_{3}(k)$ is such that
the sum of the coefficients in the $\mathrm{Q}^{A}$ expansion of each
of the building blocks of the cluster coefficients in {(\ref{oula})} gives
a counting identical to the Hofstadter case, that is $\binom{2n}{n}$.
Together with the counting $\binom{3n}{n}$ from the $c(l_{1},l_{2},
\ldots ,l_{j})$'s in {(\ref{33})} this gives the total number of closed
walks of length $3n$
\begin{equation}
\binom{3n}{n}\binom{2n}{n}=\bigg ({3n \atop n,\; n,\;n}\bigg )=
\bigg (
{\mathbf{n} \atop {\mathbf{n}}/3,\; \mathbf{n}/3,\;\mathbf{n}/3}
\bigg )
\nonumber
\end{equation}
as can also be derived from combinatorial considerations.

The algebraic area enumeration follows from rewriting the cluster
coefficients $b(n)$ as linear combinations of $\mathrm{Q}^{(1+a) A
_{u}} \, \mathrm{Q}^{(1-a) A_{d}} = \mathrm{Q}^{A_{u} + A_{d} +a (A
_{u} - A_{d})}$, where $A_{u}$ and $A_{d}$ are the algebraic area of the
enclosed up-vertex and down-vertex triangular cells respectively. In the symmetric
case $a=0$, where only the total area $A_{u} + A_{d}$ is counted, we get
%
\begin{equation}
b(1)=2 q \cos \big (2 \pi p/q\big ), \;
b(2)= -q \big (6 + 7 \cos (4
\pi p/q) + 2 \cos (8 \pi p/q)\big ),\; \mathrm{etc.} \nonumber
\end{equation}
Since the number $C_{\mathbf{n}}$ of walks of length \textbf{n} enclosing an algebraic
area $A$ derives, up to sign, from $\sum_A C_{\mathbf{n}} (A) \mathrm{Q}^A=(-1)^{\mathbf{n}/3-1}\,
\mathbf{n}\, b(\mathbf{n}/3)/q$ it follows that
\begin{itemize}%
\item
$C_{3}(1) = C_{3}(-1) = 3$; that is, among the $6$ closed walks of
length $3$, $3$ enclose an area $+1$ ($1$ up-vertex triangular cell)
and $3$ enclose an area $-1$ ($1$ down-vertex triangular cell) (see
{Fig.~\ref{fig2}});
\item
$C_{6}(0)=36, C_{6}(2) = C_{6}(-2) = 21, C_{6}(4) = C_{6}(-4) = 6$; that
is, among the $90$ closed walks of length $6$, $36$ enclose an area
$0$, $21$ an area $+2$ and $21$ an area $-2$, and $6$ an area $+4$ and
$6$ an area $-4$ (see {Fig.~\ref{fig3}});
\item
etc.
\end{itemize}

We postpone the full analysis of the properties, statistics and complete
algebraic enumeration $C_{\mathbf{n}}(A_{u} , A_{d} )$ of the chiral
triangular walks for a future publication.

\subsection{Examples of lattices for higher $g$}%
\label{sec6.3}
We conclude by briefly discussing an example of a $g=4$ lattice model.
The starting Hamiltonian is
\begin{equation}
\nonumber
H_{4} = i (-u^{2} + u^{-1} )\, v - i Q^{2a} v^{-3}
\end{equation}
with $a$ a real number. Again, $a$ and $\mathrm{Q}$ are the only relevant parameters in the
Hamiltonian.

The above is clearly a $g=4$ model with spectral parameters
\begin{equation}
\nonumber
{\mathbf{f}}_{4} (k) = -i \mathrm{Q}^{2k} + i \mathrm{Q}^{-k} = 2 \, e
^{i\pi k p/q} \,\sin ({3\pi k p/q}) ~,
~~~
{\mathbf{g}}_{4} (k) = -i\, e^{4i\pi a p /q}
\end{equation}
and spectral function
%
\begin{equation}
s_{4} (k) = \mathbf{g}_{4}(k) \, \mathbf{f}_{4} (k) \,\mathbf{f}_{4} (k+1)\,
\mathbf{f}_{4} (k+2)~.
\label{s4}%
\end{equation}
The lattice walks extracted from $H_{4}$ by interpreting $u$ and $v$ as
hops on the square lattice, however, are rather unnatural. Defining, now,
the operators
\begin{equation}
\nonumber
U = i \, \mathrm{Q}^{1+a} \; v^{-2} u^{-1} ~,
~~~
V = -i \, \mathrm{Q}^{a}\, v^{-1} u ~,
~~~
V U = \mathrm{Q}^{3}\, U V
\end{equation}
the Hamiltonian takes the form
%
\begin{equation}
\label{disdonc}
H_{4} = i( -U + U^{-1} ) V + \mathrm{Q}^{a} \, V^{-1}
\end{equation}
which has a more natural interpretation on the square lattice as hops
down, up-right and up-left by single steps. For generic $a$ it is also
chiral, as it assigns different areas to different triangular
half-cells, namely, $\frac{3}{2}+a$ for triangular half-cells with right angle in
the first and fourth quadrants and $\frac{3}{2}-a$ for mirror-image
triangular half-cells with right angle in the second and third quadrants.

Interestingly, the form {(\ref{disdonc})} of the Hamiltonian makes it now
a $g=2$ Hamiltonian with spectral parameters (taking into account the transition from $\mathrm Q$ to $\mathrm{Q}^3$ in $U$ and $V$)
%
\begin{equation}
{\mathbf{f}}_{2} (k) = -i \mathrm{Q}^{3k} + i \mathrm{Q}^{-3k} ~,
~~~
{\mathbf{g}}_{2} (k) = \mathrm{Q}^{a}
\nonumber \end{equation}
and spectral function
%
\begin{equation}
s_{2} (k) = 2\, e^{2i\pi a p/q} \sin ({6\pi k p/q}) ~.
\label{s2}%
\end{equation}
We therefore obtain another example of ``superfermionization,''
specifically, a mapping of a $g=2$ system with spectrum implied by
{(\ref{s2})} to a $g=4$ system with spectrum implied by {(\ref{s4})}. An
indication that the $g=2$ system is secretly a $g=4$ one is the fact
that all odd-particle number partition functions of the $g=2$ system
vanish, as can be deduced from the form of $s_{2} (k)$, or from the fact
that the corresponding closed lattice walks must have $2n$ steps down,
$n$ steps up-right and $n$ steps up-left, for a total of $\mathbf{n}=4n$
steps. So the number of $g=2$ particles is $\mathbf{n}/g = 2n$, which
is always even.

The above gives a flavor of the possibilities for higher $g$. The
identification of interesting random walks corresponding to other values
of $g$ and the pattern of equivalences between models with different
$g$ is left for future work.

\section{Conclusions and outlook}%
\label{sec7}
In conclusion, we demonstrated the equivalence between specific lattice
walk models and particles with (integer) exclusion statistics and
presented some examples of walks, beyond the square lattice ones, where this equivalence holds. We also obtained the exact
microscopic cluster coefficients for exclusion $g$ statistics.

Clearly there are several open questions and directions for further
investigations. The full treatment of the $g=3$ chiral triangular
lattice model statistics and its algebraic enumeration $C_{\mathbf{n}}(A
_{u} , A_{d} )$ is an immediate problem that will be addressed in an
upcoming publication. Other useful models of physical relevance
realizing higher-$g$ statistics should also be sought and analyzed.

The lattice walk--exclusion statistics connection is somewhat
mysterious. It would be useful to have a more intuitive understanding
of this correspondence (if one can be had). Further, a general
classification of the type of lattice walk models that admit an
exclusion statistics treatment would be highly desirable, especially in
the context of elucidating any geometric or topological properties of
such walks. Such a classification would also clarify and organize the
uncovered superfermionization phenomena, presumably revealing them as
a set of dualities between lattice models. This would also demonstrate
whether physically relevant walks, such as walks on the hexagonal
lattice, admit a formulation that connects them to exclusion statistics.

It is also tempting to speculate about possible lattice walk
realizations of statistics with fractional exclusion parameter $g$. A
formulation departing from the matrix model presented here would be
appropriate for such generalizations. The case $g=1$ degenerates into one-dimensional closed walks, but the question of whether
nontrivial walks corresponding to bosonic ($g=0$) statistics exist is also an
interesting one.

Finally, the ``holy grail'' of this quest would be the uncovering of
random walks in higher dimensional lattices that admit a statistical
interpretation. Generalizations of exclusion statistics to other types
of counting, or to higher dimensional objects (strings etc.), might
become relevant and necessary.

\vskip 0.4cm
\noindent
{{\it Acknowledgments}:}
A.P. acknowledges the support of {NSF} under grant {1519449} and the
hospitality of LPTMS at Universit\'{e} Paris-Sud (Orsay), where this
work was initiated. A.P. also benefited from an {``Aide Investissements d'Avenir'' LabEx PALM} grant ({ANR-10-LABX-0039-PALM}).

%
%


\begin{thebibliography}{}

\bibitem{nous}  S. Ouvry and S. Wu, ``The algebraic area of closed lattice random walks", Journal of Physics A: Mathematical and Theoretical, Volume 52, Number 25 (2019).

\bibitem{Hofstadter} D.R. Hofstadter, ``Energy levels and wave functions of Bloch electrons in rational and irrational magnetic fields", Phys. Rev. B {\bf 14} (1976) 2239.


\bibitem{Kreft} C. Kreft, ``Explicit Computation of the Discriminant for the Harper Equation with Rational Flux", SFB 288 Preprint No. 89 (1993).

\bibitem{Haldane} F.D.M. Haldane, ``Fractional statistics in arbitrary dimensions: A generalization of the Pauli principle", Phys. Rev. Lett. {\bf 67} (1991) 937–940; see also Y.S. Wu, ``Statistical distribution for generalized ideal gas of fractional-statistics particles", Phys. Rev. Lett. 73 (1994) 922–925.

\bibitem{NRBos} A.P. Polychronakos, ``Nonrelativistic Bosonization and Fractional Statistics," Nucl. Phys. B {\bf 324} (1989) 597.

\bibitem{Das} A.~Dasni\`eres de Veigy and S.~Ouvry, ``Equation of State of an Anyon gas in a Strong Magnetic Field", Phys. Rev. Lett. {\bf 72} (1994) 600; ``One-dimensional Statistical Mechanics for Identical Particles: the Calogero and Anyon Cases",  Mod. Phys. Lett. A 10 (1995) 1; Mod. Phys. Lett. B 9 (1995) 271.

\bibitem{Poly} A.P.~Polychronakos, ``Probabilities and path-integral realization of exclusion statistics," Phys. Lett. B {\bf 365} (1996) 202.
 

\bibitem{Ouv} For relevant reviews of the anyon model, exclusion statistics and related topics  see A.P. Polychronakos, ``Generalized statistics in one dimension," Les Houches LXIX Summer School ``Topological aspects of low dimensional systems” (1998) 415–472, 	arXiv:hep-th/9902157; S. Ouvry, ``Anyons and lowest Landau level Anyons," S\'eminaire Poincar\' e ``Le Spin!"(2007), Birkhauser Verlag AG, arXiv:0712.2174. 

\end{thebibliography}
\end{document}